\documentstyle[prb,twocolumn,aps]{revtex}

\begin{document}

\title{ \bf Two Definitions of Superfluid Density}

\author{
Nikolai V. Prokof'ev and Boris V. Svistunov}

\address{Russian Research Center
``Kurchatov Institute", 123182 Moscow, Russia }

\maketitle

\begin{abstract}
We point out that two different definitions of the superfluid
density - through {\it statistical} response to static gauge
phase and through {\it dynamic} response to altering gauge
phase - yield, generally speaking, different quantities in
$d<3$. The physics leading to this difference is
associated with the equilibrium statistics of supercurrent
states. Some experimentally observable consequences
of this fact are discussed.
\end{abstract}

\bigskip

In this report we would like to discuss the relation between
two alternative definitions of the superfluid density.
Consider a system with ring geometry. One way to define the superfluid
density is to consider system response to the gauge field generated
by the infinitesimal flux $\phi $ threading the ring
(we use units such that  $\hbar =1$ and particle mass $m=1$)
\begin{equation}
\delta {\cal F} = {\rho_S^{(F)} \over 2} {\varphi^2 \over L^{2-d}} \;,
\label{def-1}
\end{equation}
where ${\cal F}$ is the free-energy, $L$ is the system size, $d$
is the system dimensionality (for simplicity we assume that the
system size is the same in all directions - other examples are
discussed at the end of the paper), and $\varphi = 2\pi (\phi
/\phi_0)$ , where $\phi_0$ is the flux quantum. We do not
specify here what is the value of the flux quantum, since the
gauge field considered is not necessarily of elecrtro-magnetic
origin. For neutral systems (helium, ultracold atomic gases) it
can be introduced by rotating the system. More generally, we may
ascribe to the particles fictitious charge and couple it to
fictitious gauge field - the idea is to look at the response of
the system to the twisted boundary conditions on the ring (we
choose axis $\hat{\bf x}$ as the direction around the ring): $
\Psi ( \dots, x_j=L, \dots ) = e^{i\varphi }
\Psi ( \dots, x_j=0, \dots )$ [here $x_j$ is the $x$-coordinate of the 
$j$-th particle]. In fact, this relation
may be considered as our initial definition of the phase $\varphi $. The 
equilibrium particle number current is given then by
\begin{equation}
J^{(F)} = \rho_S^{(F)} \varphi /L \;.
\label{curr-1}
\end{equation}

On another hand, one may define superfluid density as a coefficient
in effective long-wavelength action which governs phase fluctuations
\cite{popov}:
\begin{equation}
F_{\mbox{\scriptsize eff}} = {\rho_S^{ } \over 2} \int dV \, 
(\nabla \Phi )^2 \;.
\label{def-2}
\end{equation}
We do not intend to discuss here under what conditions it is
possible to introduce phase field in the long-wavelength limit -
this is rather well known and studied problem (see, for example,
Refs.~\onlinecite{popov,borya}). In most general terms,
Eq.~(\ref{def-2}) works when $\nabla \Phi $ is a well defined
operator.  In dimensions $d > 1$ this would require the
existence of the topological long-range order. In 1D, where the
concept of the long-range order at any finite temperature is
meaningless, Eq.~(\ref{def-2}) is valid provided the low-energy
spectrum of the system is gapless (metallic state) and
relaxation times of current states, $1/\tau $, are very long. In
particular, if we label current states in 1D by index $I$, where
\begin{equation}
I={1 \over 2\pi} \oint dx \, \nabla \Phi = \mbox{\it integer} \; ,
\label{def-I}
\end{equation}
then the sufficient condition reads $1/\tau_I \ll  2\pi c
/(LK)$, where $c$ is the sound velocity, and $K/2$ is the
one-particle density matrix index, $\rho (x) \sim x^{-K/2}$ at
zero temperature. As we discuss below this condition is
satisfied at low temperatures, including temperatures exceeding
finite-size quantization limit $T \gg 2\pi c/L$.

Both definitions are used in literature with the unquestioned
assumption that $\rho_S^{(F)}=\rho_S^{ }$. 
In what follows, we demonstrate
that although in 3D one indeed may, with macroscopic accuracy,
make no distinction between $\rho_S^{(F)}$ and $\rho_S^{ }$, in
2D and especially in 1D, at finite temperature 
$\rho_S^{(F)} \neq \rho_S^{ }$. We then discuss some interesting
consequences associated with this inequality, and how these may be
directly tested experimentally.

We start from the well-known relation between $\rho_S^{(F)}$ and
the statistics of worldline winding numbers, $M$, 
(see Ref.~\onlinecite{ceperley})
\begin{equation}
\rho_S^{(F)} = {T \over L^{d-2}} \, \langle M^2 \rangle \; ,
\label{rSF-W}
\end{equation}
where $M$ is integer. Formally, Eq.~(\ref{rSF-W}) immediately
follows from the $2\pi$-periodicity in $\varphi$ of the
partition function $Z(\varphi)$, implying that it can be
Fourier-expanded as $Z(\varphi)= \sum_{M=-\infty}^{\infty}
Z_M e^{iM\varphi}$.  The statistical average in
Eq.~(\ref{rSF-W}) is taken with the distribution function $Z_M$.
The actual extreme usefulness of these formulae for numerical
simulations comes from the fact that the numbers $M$ are
directly related to the topology of the closed particle
worldlines in imaginary time. \cite{ceperley}

On the other hand, the partition function $Z(\varphi)$ can be
calculated directly from the effective long-range action
(\ref{def-2}), employing gauge-invariance argument that external 
vector-potential and/or twisted boundary condition may be
transformed into the phase of the wavefunction. We thus write
\begin{equation}
Z(\varphi ) \propto \int {\cal D} \Phi \, 
\exp \{ -F_{\mbox {\scriptsize eff}}[\Phi ] \} \; ,
\label{Z}
\end{equation}
where $\Phi$ is parameterized as
\begin{equation}
\Phi = \Phi_0 + {x \over L} (2\pi I - \varphi) \;
\label{Phi}
\end{equation}
\begin{equation}
\oint d{\bf l} \, \nabla \Phi_0 = 0 \; ,
\label{cond}
\end{equation}
\begin{equation}
{\cal D}\Phi \, \to \sum_{I=-\infty}^{\infty}{\cal D}\Phi_0 \; .
\label{DPhi}
\end{equation}
We then immediately see that
\begin{equation}
Z(\varphi) = Z_0 \sum_{I=-\infty}^{\infty} 
\exp \left\{ - { \rho_S^{ } L^{d-2} \over 2T} (2\pi I - \varphi)^2 
\right\} \; ,
\label{Zphi}
\end{equation}
where $Z_0$ is independent of $\varphi$. 

The second derivative of $-T\ln Z(\varphi)$ at $\varphi=0$,
related to $\rho_S^{(F)} L^{d-2}$, then yields
\begin{equation}
\rho_S^{(F)} = \rho_S^{ } \left( 1- {4\pi^2 \rho_S^{ } L^{d-2} \over T} \, 
\langle I^2 \rangle \right) \; ,
\label{main-1}
\end{equation}
the averaging being performed with the normalized distribution
\begin{equation}
W_I \propto 
\exp \left\{ -{2\pi^2 \rho_S^{ } L^{d-2} \over T } \, I^2 \right\} \; .
\label{distr-I}
\end{equation}
Eqs.~(\ref{main-1}) and (\ref{distr-I}) establish a relation between
$\rho_S^{(F)}$ and $\rho_S^{ }$. One can also directly relate
$\rho_S^{ }$ to the statistics of worldline winding numbers.
Using the known properties of $\theta_3$-functions, the following 
identity is valid
\begin{eqnarray}
Z(\varphi ) & \propto & \sum_{I=-\infty }^{\infty} \exp \left\{
- {\rho_S^{ } L^{d-2} \over 2T}  (2\pi I -\varphi )^2 \right\} \nonumber \\
& \equiv & 
\sum_{M=-\infty }^{\infty} \exp \left\{
- {T \over 2 \rho_S^{ } L^{d-2}}M^2 - iM \varphi \right\} \; , 
\label{another}
\end{eqnarray}
where $M$'s are some integers, which, by the uniqueness of the Fourier
transform, are immediately identified with the worldline winding numbers.
We thus see that the worldline winding numbers have Gaussian
distribution
\begin{equation}
Z_M \propto \exp \left\{
- {T \over 2 \rho_S^{ } L^{d-2}} M^2 \right\} \; .
\label{distr-M}
\end{equation}

As is clear from Eq.~(\ref{main-1}), $\rho_S^{(F)}$ and $\rho_S^{ }$
in general differ. However, at $T \to 0$ and finite $L$,
i.e. in the ``mesoscopic" limit, we find that $\rho_S^{(F)} = \rho_S^{ }$.
Moreover, in 3D case, one may safely replace $\rho_S^{(F)}$ with $\rho_S^{ }$
since, according to Eqs.~(\ref{main-1}-\ref{distr-I}), finite-size
corrections are exponentially small in macroscopic parameter:
the probability of finding non-zero $I= \pm 1$ dies away
as $\exp \{ -2\pi^2 \rho_S^{ } L/T \}$. This is probably the reason why
it became a custom to identify $\rho_S^{(F)}$ and $\rho_S^{ }$.

In 1D systems the situation changes drastically. In the
thermodynamic limit, when $L \to \infty$ first, that is $TL \to
\infty$ at any finite temperature, we have negligible
$\rho_S^{(F)}$ [$\sim LT \exp \{ -LT/2\rho_S^{ } \}$, see
Eq.~(\ref{distr-M})] while $\rho_S^{ }$ is finite.
At this point we would like to emphasize that zero $\rho_S^{(F)}$
is {\it not} in contradiction with the fact that the system may
remain superfluid in any dynamic sense. Equations 
(\ref{main-1}-\ref{distr-I}) make it clear that $\rho_S^{(F)} \to 0$
simply due to the broad distribution over the current states,
with typical $I \sim (LT)^{1/2} \gg 1$. However, relaxation times
of current states in 1D may be very long \cite{our} (see also below),
up to astronomical scale. Therefore, out-of-equilibrium state
obtained experimentally, say, by switching on/off the gauge
phase $\varphi$, will support a supercurrent $j \sim \rho_S^{ } \varphi /L$.

To estimate relaxation times $\tau_I$ in 1D rings (obviously,
for the relaxation time to be finite the translational
invariance must be violated by impurities, or lattice potential)
one may use an effective Hamiltonian describing low energy
dynamics of supercurrent states and density fluctuations \cite{our}.
Relaxation rate is obtained then from the golden-rule expression
for the terms breaking translational invariance. Formally, this problem
is identical to the study of particle dynamics with Ohmic
coupling to the oscillator bath environment \cite{caldeira}
- an analogy widely used in 1D systems since Ref.~\onlinecite{kane}.
The standard result for the transition rate from $I$ to $I+1$ due to
single impurity backscattering potential, $\Delta $, is given by
(see, e.g., Ref.~\onlinecite{reviews})
\begin{equation}
\tau^{-1}_I \sim {\Delta^2 \over \omega_o} 
\left({T \over \omega_o } \right)^{2/K-1} \;,
\label{1imp}
\end{equation}
where $\omega_o$ is some model-dependent high-energy cut-off. If
parameter $K$ is small, then at low temperature one may easily
get extremely long $\tau_I$. The answer for the disordered ring
with white-noise disorder is just the single impurity expression 
(\ref{1imp}) multiplied by $L$.

One may directly test the striking difference between
$\rho_S^{(F)}$ and $\rho_S^{ }$ in 1D systems experimentally. In
fact, we propose a unique measuring device for magnetic fields,
free of limitations set by quantization of magnetic flux.
Suppose we would like to measure local magnetic field at
the point ${\bf R}$ by placing at this point a device recording
magnetic flux through some small surface with subsequent reading
the information recorded. We point out that such a device can
be realized as an ensemble of 1D (quasi-)superconducting rings.
Being placed at the point ${\bf R}$ at high temperature 
(say, $T \sim \omega_0$) when relaxation times are short, and then cooled
down to low temperatures when relaxation times become very long
(much longer than expected ``storage time"), the ensemble will not
respond to the local field, since in this case the response is
{\it statistical} and $\rho_S^{(F)} \approx 0$. However, removing
cooled device from the point ${\bf R}$ we excite finite supercurrent,
since now we probe {\it dynamic} response $\rho_S^{ }$.
It is important to note, that, in contrast to the case of massive
superconducting ring, the stored magnetic flux is not quantized.

Another interesting experiment directly follows from Eq.~(\ref{distr-I}).
A single 1D ring being cooled down to low temperature will typically end up in
a state with non-zero supercurrent $j \sim (\rho_S^{ } 
T_{\mbox{\scriptsize ex}}/L)^{1/2}$, where $T_{\mbox{\scriptsize ex}}$
is the ``freezing" temperature at which the current relaxation
time becomes comparable to the experimental cooling time.

The 2D case is even more intriguing. For an isotropic system with
$L_x=L_y$, Eqs.~(\ref{main-1}), (\ref{distr-I}) and
(\ref{distr-M}) are system-size independent, in particular
\begin{eqnarray}
\rho_S^{(F)} & = & \rho_S^{ } \left( 1- {4\pi^2 \rho_S^{ } \over T} \, 
\langle I^2 \rangle \right) \; , \nonumber \\
W_I & \propto & 
\exp \left\{ -{2\pi^2 \rho_S^{ } \over T } \, I^2 \right\} \; .
\label{2dcase}
\end{eqnarray}
The famous Nelson-Kosterlitz formula
\begin{equation}
T_c={\pi \over 2} \rho_S^{ }
\label{NK}
\end{equation}
relates transition temperature and $\rho_S^{ }$ [not $\rho_S^{(F)}$!]
(see Ref.~\onlinecite{nelson}, and also Ref.~\onlinecite{borya}). 
Substituting it into Eq.~(\ref{2dcase}), we find that
with very high accuracy
\begin{equation}
\rho_S^{(F)}(T_c) = \rho_S^{ } ( 1- 16 \pi e^{-4\pi} ) \; . 
\label{corr}
\end{equation}
For purely numerical reasons the relative difference
$(\rho_S^{(F)} - \rho_S^{ })/\rho_S^{ } < 2 \cdot 10^{-4}$
is very small. That is why in a recent state-of-the-art
numerical study of the Berezinskii-Kosterlitz-Thouless
transition \cite{kawashima} formally incorrect relation
$T_c=(\pi / 2) \rho_S^{(F)}$ 
(or $\langle M^2 \rangle =2/\pi$ \cite{ceperley})
was successfully used to fit the data. 
However, unlike in 3D case, the difference
between $\rho_S^{(F)}$ and $\rho_S^{ }$ is not controlled any more
by small parameters. Furthermore, if a 2D system is
elongated geometrically: $L_x > L_y$, then
\begin{eqnarray}
\rho_{S,x}^{(F)} & =& \rho_S^{ } \left( 1- {4\pi^2 \rho_S^{ } L_y \over T L_x} 
\, 
\langle I^2 \rangle \right) \; ,  \nonumber \\
W_{I,x} & \propto &
\exp \left\{ -{2\pi^2 \rho_S^{ } L_y \over T L_x } \, I^2 \right\} \; ,
\label{2dcase-2}
\end{eqnarray}
where index $x$ for $\rho_S^{(F)}$ and $W_I$ is used to specify
that we are dealing with the response to the twisted boundary
condition along $x$-axis. Clearly, at $L_x \gg L_y$ the physics
becomes analogous to that of 1D rings. We thus see that in 2D $\rho^{F}_S$
depends on the sample geometry and at $L_x \geq 4\pi  L_y$,
the value of $\rho_S^{(F)}$ is not related to $T_c$,
Eq.~(\ref{NK}), even approximately.

It seems rather easy experimentally to confirm these conclusions
by producing narrow Carbino-disk samples. Since in this case we
are dealing with macroscopic phase transition phenomenon, at $T
< T_c$ the current relaxation times become astronomically long
and the above-mentioned experiments on 1D rings would simply
require cooling below $T_c$.

We are grateful to Yu.~Kagan for his interest and discussions.
This work was supported by the Russian Foundation for Basic
Research (under Grant No. 98-02-16262) and by the Grant 
INTAS-97-0972 [of the European Community].


\begin{thebibliography}{99}
\bibitem{popov} V.N. Popov, {\it Functional Integrals in Quantum 
Field Theory and Statistical Physics}, Dordrecht: Reidel (1983). 

\bibitem{borya} B.V. Svistunov, 
                J. Moscow Phys. Soc. {\bf 2}, 283 (1992).

\bibitem{ceperley} E.L. Pollock and D.M. Ceperley, 
                  Phys. Rev. {\bf B 36}, 8343 (1987).

\bibitem{our} V.A. Kashurnikov, A.I. Podlivaev, N.V. Prokof'ev, 
              and B.V. Svistunov,
             Phys. Rev. {\bf B 53}, 13091 (1996).

\bibitem{caldeira} A.O. Caldeira and A.J. Leggett, 
                   Ann. Phys. {\bf 149}, 374 (1983).

\bibitem{kane} C.L. Kane and M.P.A. Fisher, 
               Phys. Rev. Lett. {\bf 68}, 1220 (1992).

\bibitem{reviews}  A.J. Leggett, S. Chakravarty, A.T. Dorsey, M.P.A.
                   Fisher, A. Garg, and W. Zwerger,
                   Rev. Mod. Phys. {\bf 59}, 1 (1987);
                   {\it Quantum Tunneling in Condensed Media},
                   eds. Yu. Kagan and A.J. Leggett, North-Holland,
                   Elsevier (1992). 

\bibitem{nelson} D.R. Nelson and J.M. Kosterlitz, Phys. Rev. Lett. 
                 {\bf 39} 1201 (1977).

\bibitem{kawashima} K. Harada and N. Kawashima, J. Phys. Soc. Jpn.
                    {\bf 67}, 2768 (1998).
\end{thebibliography}
\end{document}